\documentclass[sigconf]{acmart}
\usepackage{graphicx}
\usepackage{algorithm}
\usepackage{rotating}
\usepackage{balance}

\usepackage{tabularx}
\usepackage{subfigure}
\usepackage{bm}
\usepackage{multirow}
\usepackage{enumitem}

\AtBeginDocument{%
  }

\copyrightyear{2024}
\acmYear{2024}
\setcopyright{acmlicensed}
\acmConference[SIGIR '24]{Proceedings of the 47th International ACM SIGIR Conference on Research and Development in Information Retrieval}{July 14--18, 2024}{Washington, DC, USA}
\acmBooktitle{Proceedings of the 47th International ACM SIGIR Conference on Research and Development in Information Retrieval (SIGIR '24), July 14--18, 2024, Washington, DC, USA}
\acmDOI{10.1145/3626772.3661369}
\acmISBN{979-8-4007-0431-4/24/07}

\begin{document}

\title{Interest Clock: Time Perception in Real-Time Streaming Recommendation System}

\author{Yongchun Zhu}
\affiliation{%
  \institution{Douyin Group}
  \city{Beijing}
  \country{China}}
\email{zhuyc0204@gmail.com}

\author{Jingwu Chen$^{\dag}$}
\affiliation{%
  \institution{Douyin Group}
  \city{Beijing}
  \country{China}}
\email{chenjingwu@bytedance.com}

\author{Ling Chen}
\affiliation{%
  \institution{Douyin Group}
  \city{Shanghai}
  \country{China}}
\email{chenling.nb@bytedance.com}

\author{Yitan Li}
\affiliation{%
  \institution{Douyin Group}
  \city{Shanghai}
  \country{China}}
\email{liyitan@bytedance.com}

\author{Feng Zhang}
\affiliation{%
  \institution{Douyin Group}
  \city{Shanghai}
  \country{China}}
\email{feng.zhang@bytedance.com}

\author{Zuotao Liu}
\affiliation{%
  \institution{Douyin Group}
  \city{Shanghai}
  \country{China}}
\email{michael.liu@bytedance.com}

\thanks{
$\dag$ Jingwu Chen is the corresponding author.}

\renewcommand{\shortauthors}{Zhu, et al.}

\begin{abstract}
User preferences follow a dynamic pattern over a day, e.g., at 8 am, a user might prefer to read news, while at 8 pm, they might prefer to watch movies. Time modeling aims to enable recommendation systems to perceive time changes to capture users' dynamic preferences over time, which is an important and challenging problem in recommendation systems. Especially, streaming recommendation systems in the industry, with only available samples of the current moment, present greater challenges for time modeling. There is still a lack of effective time modeling methods for streaming recommendation systems.
In this paper, we propose an effective and universal method Interest Clock to perceive time information in recommendation systems. 
Interest Clock first encodes users' time-aware preferences into a clock (hour-level personalized features) and then uses Gaussian distribution to smooth and aggregate them into the final interest clock embedding according to the current time for the final prediction. 
By arming base models with Interest Clock, we conduct online A/B tests, obtaining +0.509\% and +0.758\% improvements on user active days and app duration respectively. Besides, the extended offline experiments show improvements as well. Interest Clock has been deployed on Douyin Music App. 
\end{abstract}




\maketitle

\section{Introduction}\label{sec:1}
Generally, user interests vary from user to user, named personalized user preferences. Most existing recommendation methods~\cite{davidson2010youtube,covington2016deep,zhang2016collaborative,zhou2018deep,bhagat2018buy,wang2021dcn} focus on modeling static personalized preferences. However, these methods overlook the fact that users' preferences are dynamic and fluctuate with time. For example, in a short video platform, users could prefer news videos in the morning, while they like entertainment videos at night. In a music platform, users like listening to DJs in the morning and sleep-inducing music at night. Thus, it is important to enable recommendation models to perceive time information to provide time-aware personalized service for users, which could significantly improve user experiences.

Time perception in recommendation is a very challenging problem, and only a few works attempt to address this problem. For takeaway
recommendation, \citet{zhang2023modeling} divided a day into four periods, including morning, noon, night, and last night, and used different graph models for different periods. However, takeaway recommendation inherently has period differences, while other recommendation systems not, e.g., short video platforms and music platforms. In addition, some practical methods encode time gap information in sequential methods~\cite{tang2018personalized,zhou2018deep,pi2020search,chang2023twin}, which can guide the importance learning of sequential information. However, these sequential methods ignore dynamic preferences over time. To perceive time information, a widely adopted method in the industry is encoding the hour of a day and the day of a week into hour embeddings and day embeddings, named time encoding~\cite{ping2021user,li2022automatically}, which achieves remarkable performance.

The early recommendation systems adopt a daily training framework, collecting all samples of one day, and shuffling them for training. The time encoding methods work well in the daily training framework. However, in recent years, to improve the timeliness of recommendation systems, many platforms have upgraded the daily training framework to a real-time streaming training framework, in which samples are used for training immediately after they are produced. The real-time streaming framework proposes a new challenge for time perception. In the streaming framework, all training samples at a certain moment have the same time features, and recommendation systems are capable of producing tens of millions of samples every hour, which leads to the recommendation model only fitting current time features and forgetting other time information. This discreteness of the time encoding methods can result in periodical online patterns and introduce instability, which cannot work well in streaming recommendation systems.

In this paper, we propose an effective and universal method Interest Clock to perceive time information in streaming recommendation systems. The key idea of the proposed method is encoding personalized user interests of 24 hours into a clock. Firstly, we calculate the user's past interests by hour and store the time-aware features in samples. The time-aware features are discrete, with one feature corresponding to each hour. However, users' interests do not change abruptly, e.g., it is unlikely for a user's interests to be significantly different between 7:59 and 8:01. To address the issue of interest abrupt changes caused by discrete interest clock features, we use empirical Gaussian distribution to smooth and aggregate the interest clock features of 24 hours. The proposed Interest Clock method transforms time modeling into time-aware feature modeling. For a certain moment, different users have various time-aware preference embeddings, which can cover the overall feature space. Thus, the proposed method can solve the periodical online pattern and instability problems of time encoding methods in real-time streaming recommendation systems. 

The main contributions of our work are summarized in three folds:
\begin{itemize}
    \item To enable recommendation systems to perceive time information, we propose an effective and universal method named Interest Clock. To the best of our knowledge, we are the first to tackle the time perception problem in real-time streaming recommendation systems.
    \item We conduct online experiments, obtaining +0.509\% and +0.758\% improvements on user active days and app duration respectively, which obtains the biggest improvement of a single model in 2023. In addition, offline experiments also demonstrate its effectiveness.
    \item Interest Clock has been widely deployed in online recommendation systems of Douyin Music App, indicating its superior effectiveness and universality.
\end{itemize}

\section{Related Work}
Time encoding~\cite{ping2021user,li2022automatically} is a widely adopted method in the industry to perceive time information, which encodes the hour of a day and the day of a week into hour embeddings and day embeddings. However, time encoding methods transform time into discrete embeddings, which can not work in modern real-time streaming recommendation systems. For takeaway
recommendation, \citet{zhang2023modeling} divided a day into four periods, including morning, noon, night, and last night, and used different graph models for different periods, which is difficult to deploy in other scenarios. In addition, industrial engineers usually encode time gap in sequential methods~\cite{tang2018personalized,zhou2018deep,pi2020search,chang2023twin}, which is only capable of better learning the significance of sequential information, failing to directly model time information. To the best of our knowledge, we are the first to tackle the time perception problem in real-time streaming recommendation systems.

\section{Proposed Method}

\begin{figure}[t]
	\centering
	\begin{minipage}[b]{0.95\linewidth}
		\centering
		\includegraphics[width=1.\linewidth]{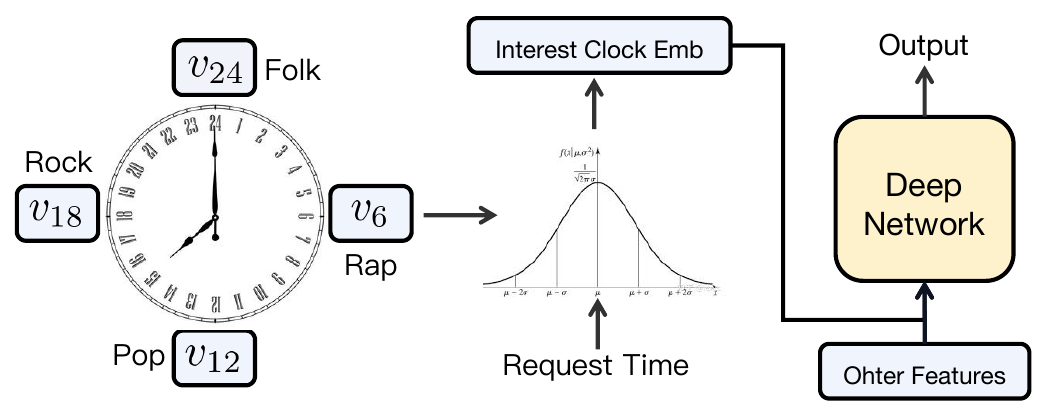}
	\end{minipage}
	\caption{Interest Clock first encodes users' time-aware preferences into a clock (hour-level personalized features) and then uses Gaussian distribution to smooth and aggregate them into the final interest clock embedding according to the current time for the final prediction.}\label{fig:model}
\end{figure}

In this section, we introduce the details of our proposed method Interest Clock. In Section~\ref{sec:3.1}, we specify the common setup of a recommendation task in industrial recommendation systems. In Section~\ref{sec:3.2}, we introduce the details of feature extraction. In Section~\ref{sec:3.3}, we introduce the proposed method.

\subsection{Recommendation Task Setup}\label{sec:3.1}
First, we consider the common setup for a binary classification task, such as CTR prediction in recommendation systems. Each sample consists of the input raw features and a label $y \in \{0,1\}$, and these features are transformed into low-dimensional representations, named feature embeddings, denoted as $[\bm{v}_1, \cdots, \bm{v}_n]$, where $n$ indicates the number of raw features. The prediction of a recommendation model $f(\cdot)$ with the embeddings as inputs is formulated as:
\begin{equation}
    \hat{y} = f([\bm{v}_1, \cdots, \bm{v}_n]).
\end{equation}

The cross-entropy loss is often used as the optimization target for binary classification:
\begin{equation}
    \mathcal{L} = -y \log \hat{y} - (1-y) \log (1 - \hat{y}).\label{eq:loss}
\end{equation}

In this paper, we focus on the representation of time information, denoted as $\bm{v}_{time}$.

\begin{table*}[htbp]
\centering
\setlength\tabcolsep{3pt}
\caption{Online A/B testing results of a ranking task. Each row indicates the relative improvement with our Interest Clock over the baseline (a DCN-V2-based multi-task model). The square brackets represent the 95\% confidence intervals for online metrics. Statistically significant improvement is marked with bold font in the table. Low-, Middle-, and High-active indicate different user groups.}
\begin{tabular}{lcccccc}
\toprule
\multirow{2}{*}{}                    & \multicolumn{2}{c}{Main Metrics}                    & \multicolumn{4}{c}{Constraint Metrics}                                                                    \\
\cmidrule(r){2-3}  \cmidrule(r){4-7}
                                     & Active Day               & Duration                 & Like                     & Finish                   & Comment                  & Play                     \\
\midrule
\multirow{2}{*}{Low-active}    & \textbf{0.731\%}         & \textbf{1.201\%}         & 1.327\%                  & \textbf{0.956\%}         & \textbf{0.831\%}         & \textbf{1.371\%}         \\
                                     & {[}-0.267\%, +0.267\%{]} & {[}-0.789\%, +0.789\%{]} & {[}-2.505\%,+2.504\%{]}  & {[}-0.337\%, +0.338\%{]} & {[}-0.606\%, +0.606\%{]} & {[}-0.527\%, +0.527\%{]} \\
\multirow{2}{*}{Middle-active} & \textbf{0.625\%}         & \textbf{1.205\%}         & 1.738\%                  & \textbf{0.684\%}         & \textbf{1.253\%}         & \textbf{1.340\%}         \\
                                     & {[}-0.151\%, +0.152\%{]} & {[}-0.429\%, +0.429\%{]} & {[}-2.243\%, +2.243\%{]} & {[}-0.166\%, +0.166\%{]} & {[}-0.456\%, +0.457\%{]} & {[}-0.293\%, +0.293\%{]} \\
\multirow{2}{*}{High-active}   & \textbf{0.437\%}         & \textbf{0.574\%}         & \textbf{2.187\%}         & \textbf{0.427\%}         & \textbf{0.833\%}         & \textbf{0.929\%}         \\
                                     & {[}-0.091\%, +0.092\%{]} & {[}-0.307\%, +0.308\%{]} & {[}-1.995\%, +1.996\%{]} & {[}-0.095\%, +0.095\%{]} & {[}-0.441\%, +0.442\%{]} & {[}-0.235\%, +0.235\%{]} \\
\midrule
\multirow{2}{*}{Overall}             & \textbf{0.509\%}         & \textbf{0.758\%}         & \textbf{1.514\%}         & \textbf{0.617\%}         & \textbf{0.962\%}         & \textbf{1.136\%}         \\
                                     & {[}-0.073\%, +0.073\%{]} & {[}-0.218\%, +0.218\%{]} & {[}-0.925\%, +0.926\%{]} & {[}-0.107\%, +0.108\%{]} & {[}-0.249\%, +0.249\%{]} & {[}-0.174\%, +0.175\%{]}\\
\bottomrule
\end{tabular}\label{tab:online}
\end{table*}

\begin{table}[htbp]
\centering
\caption{Offline results (AUC and UAUC) on the industrial datasets DouyinMusic-20B.}
\begin{tabular}{lcc}
\toprule
               & AUC    & UAUC   \\
\midrule
Baseline    & 0.6631 & 0.6007 \\
Naive Clock    & 0.6666 & 0.6015 \\
Adaptive Clock & 0.6662 & 0.5859 \\
Gaussian Clock & \textbf{0.6695} & \textbf{0.6069} \\
\bottomrule
\end{tabular}\label{tab:offline}
\end{table}

\subsection{Feature Engineer}\label{sec:3.2}

The simple time encoding methods~\cite{ping2021user,li2022automatically} directly concatenate the embeddings of the current hour $\bm{v}_{hour}$ and the embedding of the current day $\bm{v}_{day}$ into a time embedding, denoted as $\bm{v}_{time} = [\bm{v}_{hour}; \bm{v}_{day}]$. Our proposed method aims to encode time-aware personalized preferences. Thus, the first step is extracting time-aware personalized features.

Firstly, we split a day into 24 buckets to represent 24 hours of a day. Then, we compute users' time-aware preferences from the consumption data of users in a certain hour in the past 30 days. For example, we obtain all samples generated by users from 7:00 to 8:00 in the past 30 days, and each sample has multiple labels (e.g., like, skip, finish, dislike) and many features (e.g., genre, mood, language). The score of each feature is computed as:
\begin{equation}
    score_{fea} = \alpha * Cnt_{like} + \beta * Cnt_{finish} - \gamma * Cnt_{Skip} - \omega * Cnt_{dislike},\label{eq:vt}
\end{equation}
where $\alpha, \beta, \gamma, \omega$ are the hyperparameters. $Cnt$ indicates the number of samples of the corresponding behavior, and the samples contain the target feature denoted as $fea$.

With Equation~(\ref{eq:vt}), we calculate the score of a certain hour for several given features, including genre, mood, and language, and the top three genre/mood/language features are used as the time-aware features. Thus, the embeddings of the time-aware features, e.g., genre, are denoted as $\bm{v}_{time}^{genre} = [\bm{v}^{genre}_{1}, \bm{v}^{genre}_{2}, \cdots, \bm{v}^{genre}_{24}]$. Similarly, the embeddings of other time-aware features can be obtained in the same way, denoted as $\bm{v}_{time}^{mood}, \bm{v}_{time}^{lang}$.

\subsection{Interest Clock}\label{sec:3.3}
Figure~\ref{fig:model} overviews our proposed method Interest Clock, whose goal is to enable the model to perceive time information in streaming recommendation systems. With the feature extraction procedure, we have encoded users' time-aware personalized preference into a clock, i.e., the hour-level features. Two simple methods can be exploited to aggregate interest clock features, (1) concatenate the interest embeddings of 24 hours into one embedding, and (2) according to current request time $t$, feed the model with the corresponding interest embedding $\bm{v}_t$.

The first method relies on adaptive learning of the importance of each hour-level feature by an optimization procedure, called Adaptive Clock in this paper. However, we find it difficult for deep models to adaptively learn the feature weights, because the model would overfit the current time and forget information of other time in streaming systems (the same problem in time encoding methods as introduced in Section~\ref{sec:1}). The second method only uses the time-aware preference embedding of the current time, called Naive Clock in this paper. However, Naive Clock faces a sudden change of time-aware features at an hourly time.

To solve the shortcomings of the above two methods, we propose Gaussian Interest Clock, which aggregates the time-aware embeddings of 24 hours with an empirical Gaussian distribution. The interest clock embedding can be formulated as:
\begin{equation}
    \begin{split}
        \bm{v}_{clock} &= \sum_{t=1}^{24} g(\delta_{time}) [\bm{v}^{genre}_t, \bm{v}^{mood}_t, \bm{v}^{lang}_t],\\
        \delta_{time} &= \min ( \mod(t+24-cur\_time, 24), \\
        & \mod(cur\_time+24-t, 24)), \\
        g(\delta_{time}) &=  \frac{1}{\sqrt{2\pi}\sigma} \exp \left( -\frac{(\delta_{time}-\mu)^2}{2\sigma^2} \right),
    \end{split}
\end{equation}
where $\sigma, \mu$ are empirically set to $1$ and $0$, and $cur\_time$ indicates current request time. Finally, the interest clock embedding is concatenated with other feature embeddings and fed into a deep network for predictions. The overall framework is trained with the cross-entropy loss as Equation~(\ref{eq:loss}).

\section{Experiments}

\begin{figure*}[t]
	\centering
	\begin{minipage}[b]{0.8\linewidth}
		\centering
		\includegraphics[width=1.\linewidth]{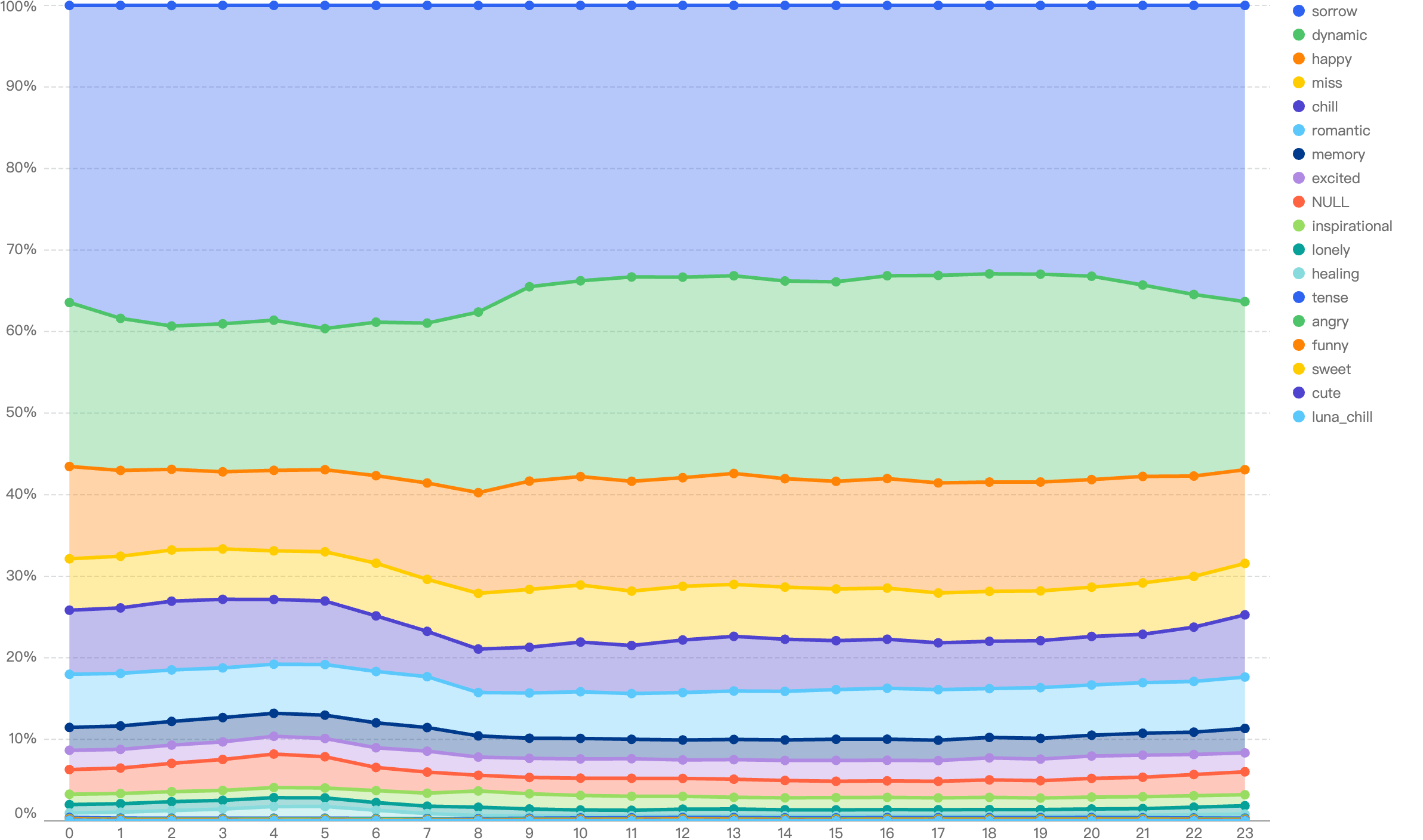}
	\end{minipage}
	\caption{Analysis of the time information in recommendation systems. The horizontal axis represents hours, and the vertical axis represents the percentage of impression counts.}\label{fig:analysis}
\end{figure*}

In this section, we conduct extensive offline and online experiments with the aim of answering the following evaluation questions: 
\begin{itemize}
     \item[\textbf{EQ1}] Can Interest Clock bring improvement to the performance of online recommendation tasks?
     \item[\textbf{EQ2}] How does the Interest Clock perform in industrial datasets?
     \item[\textbf{EQ3}] What are the effects of time information in real-world recommendation systems?
\end{itemize}

\textbf{Datasets.} We evaluate Interest Clock with baselines on a large-scale industrial recommendation dataset. 

\textit{DouyinMusic-20B}: Douyin provides a music recommendation service, with over 10 million daily active users. We collect from the impression logs and get one dataset. The dataset contains more than 20 billion samples, denoted as \textit{DouyinMusic-20B}. Each sample of the industrial datasets contains more than one hundred features, including both non-ID meta features (gender, age, genre, mood, scene, and so on) and ID-based personalized features (user ID, item ID, artist ID, interacted ID sequence), which can represent the real-world scenarios. We use `Finish' as the label. The DouyinMusic-20B dataset contains samples from Douyin Music across the time span of 8 weeks from August to September 2023. Then, we take the first 6 weeks as the training set, the following 1 week as the validation set, and the remaining 1 week as the test set. 

\textbf{Online A/B Testing (EQ1).} To verify the real benefits Interest Clock brings to our system, we conducted online A/B testing experiments for more than one month for the ranking task in Douyin Music App. We evaluate model performance based on two main metrics, Active Days and Duration. We also take additional metrics, which evaluate user engagement, including Like, Finish, Comment, and Play, which are usually used as constraint metrics. We apply the proposed Interest Clock on a DCN-V2-based multi-task model~\cite{wang2021dcn} which is deployed in the online ranking tasks. The online A/B results of low-, middle-, high-active, and whole users are shown in Table~\ref{tab:online}. For the main metrics Active Days and Duration, the proposed Interest Clock achieves a large improvement of +0.509\% and +0.758\% for all users with statistical significance, which is remarkable given the fact that the average Active Days and Duration improvement from production algorithms is around 0.05\% and 0.1\% respectively. In addition, the results demonstrate that Interest Clock could improve the recommendation performance for users of different activity levels.

\textbf{Offline Results (EQ2).} We adopt AUC and UAUC as offline metrics. We use Naive, Adaptive, and Gaussian Interest Clock to replace the time encoding methods in the online baseline DCN-V2-based multi-task model. The experimental results on the industrial dataset are shown in Table~\ref{tab:offline}. The results further reveal several insightful observations. Gaussian Interest Clock could outperform the best baseline significantly. UAUC of Adaptive Clock is worse than the baseline, and the reason could be adaptive weights of time information are difficult to learn in streaming recommendation systems. We find that Gaussian Clock outperforms Naive Clock, which demonstrates empirical Gaussian weights are effective.

\textbf{Analysis (EQ3).} To analyze the influence of time information in recommendation systems, we visualized the distribution of music mood tags at different time as shown in Figure~\ref{fig:analysis}. The results further reveal several insightful observations. (1) The distribution of content provided by recommendation systems varies over time, which demonstrates the users' preferences follow a dynamic pattern over a day. (2) The overall content distribution is consistent with our intuition. For example, the sorrow songs account for more impressions in 0:00-8:00 than 12:00-20:00.
\section{Conclusion}
In this paper, to enable recommendation systems to perceive time changes, we propose an effective method Interest Clock. 
Firstly, we encode users' time-aware preferences into a clock, obtaining hour-level personalized preference features. Then, we use Gaussian distribution to smooth and aggregate them into the final interest clock embedding, which is fed into a deep network for the final predictions.
We demonstrated the superior performance of the proposed Interest Clock in offline experiments. In addition, we conducted online A/B testing, obtaining +0.509\% and +0.758\%  improvements on user active days and app duration respectively, which demonstrates the effectiveness and universality of Interest Clock in online systems. 
Moreover, Interest Clock has been deployed on ranking tasks in multiple applications of Douyin Group.

\bibliographystyle{ACM-Reference-Format}
\bibliography{main}

\appendix

\section{Biography}

{Yongchun Zhu} is currently a researcher at Douyin Group, Beijing, China.
He received his Ph.D. degree from Institute of Computing Technology, Chinese Academy of Sciences, Beijing, China. His main research interests include recommendation systems and transfer learning. He has published over 30 papers in top-tier international conferences and journals including KDD, WWW, SIGIR, TKDE, TNNLS and so on. Homepage: \url{https://scholar.google.com.hk/citations?user=iKUIgeQAAAAJ}

\end{document}